\documentclass[12pt,a4paper]{article}
\usepackage{amssymb}
\usepackage[T2A]{fontenc}
\usepackage[cp866]{inputenc}
\usepackage[russian,english]{babel}
\usepackage{graphicx}
\usepackage[dvips]{psfrag}
\usepackage[unicode]{hyperref}
\begin{document}
\begin{center}
{\Large Nonlinear electrodynamics with the maximum allowable symmetries}\\[3mm]
{B.~P.~Kosyakov}\\[3mm]
{{\small Russian Federal Nuclear Center--VNIIEF, 
Sarov, 607188 Nizhni{\u\i} Novgorod Region, Russia;\\
Moscow Institute of Physics {\&} Technology, Dolgoprudni{\u\i}, 141700 Moscow 
Region, Russia}\\
{\tt E-mail:} 
${\rm  kosyakov.boris@gmail.com}$ 
} 
\end{center}
\begin{abstract}
\noindent
{Recently Bandos, Lechner, Sorokin, and Townsend have discovered that Maxwell's 
electrodynamics can be generalized so that the resulting nonlinear theory 
preserves both conformal invariance and SO$(2)$ duality-rotation invariance.
Their result can be derived in a simpler way. 
}
\end{abstract}
It has long been known that Maxwell's equations are invariant under both 
conformal transformations \cite{Bateman}, \cite{Cunningham} and Hodge 
duality rotations \cite{Rainich}.
Is it possible to preserve these symmetries in nonlinear modifications of 
Maxwell's theory with Lagrangians of the form ${\cal L}={\cal L}({\cal S},{\cal P})$?
Here, the arguments of ${\cal L}$ are the electromagnetic field invariants
\begin{equation}
{\cal S}=\frac12\,F_{\mu\nu}F^{\mu\nu}\,,
\quad
{\cal P}=\frac12\,F_{\mu\nu}{}^\ast\! F^{\mu\nu}\, ,
\label{EM-invariants}
\end{equation}
the field strength $F_{\mu\nu}$ is expressed in terms of vector potentials,  
\begin{equation}
F_{\mu\nu}=\partial_\mu A_\nu-\partial_\nu A_\mu\,, 
\label{EM-field-strength}
\end{equation}
and the Hodge dual of $F_{\mu\nu}$ is defined by 
\begin{equation}
{}^\ast F^{\mu\nu}=\frac12\,\epsilon^{\mu\nu\rho\sigma}F_{\rho\sigma}\,.
\label{EM-dual-field}
\end{equation}
Recently Bandos, Lechner, Sorokin, and Townsend have demonstrated \cite{Bandos} 
that such is the case.
This is a profound result.
Indeed, the analysis of this issue, apart from its cognitive significance, 
may show the utility in the low-energy effective theory to strings.

However, the line of reasoning in Ref.~\cite{Bandos} may seem somewhat meandering.
Let us obtain the same result in a direct and simpler way.   

We begin with the Bessel-Hagen criterion for conformal invariance \cite{Bessel-Hagen} 
\begin{equation}
\Theta^\mu_{~\mu} =0\,,
\label{Bessel-Hagen}
\end{equation}
where $\Theta_{\mu\nu}$ is the symmetric stress-energy tensor of electromagnetic 
field. 
Equation (\ref{Bessel-Hagen}) can be cast \cite{Kosyakov} as follows:
\begin{equation}
{\cal L}_{\cal S}\,{\cal S}+{\cal L}_{\cal P}\,{\cal P}={\cal L}\,,
\label
{energy-tensor-conv-nonl}
\end{equation}                          
where ${\cal L}_{\cal S}=\partial{\cal L}/\partial{\cal S}$ and
${\cal L}_{\cal P}=\partial{\cal L}/\partial{\cal P}$.
We then notice that the Euler--Lagrange equations 
\begin{equation}
\partial_\mu E^{\mu\nu}=0\,, 
\label
{partial-ast-G=j}
\end{equation}                          
in which the excitation $E_{\mu\nu}$ is defined by
\begin{equation}
E_{\mu\nu}=\frac{\partial{\cal L}}{\partial F^{\mu\nu}}
=2\left({\cal L}_{\cal S}\,F_{\mu\nu}+
{\cal L}_{\cal P}\,{}^\ast\!F_{\mu\nu}\right),
\label
{ast-G-df}
\end{equation}                          
and the Bianchi identity
\begin{equation}
\partial_\mu{}^\ast\!F^{\mu\nu}=0\,,
\label
{Bianchi-partial-ast-F}
\end{equation}                          
which is a mere restatement of Eq.~(\ref{EM-field-strength}), are invariant 
under the general electric-magnetic duality rotation
\begin{equation}
E'_{\mu\nu}=E_{\mu\nu}\cos\theta+{}^\ast\! F_{\mu\nu}\sin\theta,
\quad
{}^\ast\!F'_{\mu\nu}={}^\ast\! F_{\mu\nu}\cos\theta-E_{\mu\nu}\sin\theta\,.
\label
{duality-transf}
\end{equation} 
However, the constitutive equations
\begin{equation}
E^{\mu\nu}=E^{\mu\nu}(F,{}^\ast\! F)\,,
\label
{constitutive}
\end{equation}                          
stemming from (\ref{ast-G-df}), are in general devoid of this invariance.                         
The Gaillard--Zumino criterion \cite{GaillardZumino} for invariance under 
the duality transformations (\ref{duality-transf}) reads
\begin{equation}
{}^\ast\! E_{\mu\nu}\,E^{\mu\nu}={}^\ast\!F_{\mu\nu}\,F^{\mu\nu}\,.
\label
{duality-cond}
\end{equation}                         

We use Eq.~(\ref{ast-G-df}) and the fact that 
${}^\ast\!F_{\mu\nu}\,{}^\ast\!F^{\mu\nu}=-F_{\mu\nu}\,F^{\mu\nu}$ to bring 
Eq.~(\ref{duality-cond}) to the form
\begin{equation}
4\left({\cal L}_{\cal S}^2-{\cal L}_{\cal P}^2\right){\cal P}
-8\,{\cal L}_{\cal S}\,{\cal L}_{\cal P}\,{\cal S}={\cal P}\,.
\label
{Gaillard-Zumino-eq}
\end{equation} 
We multiply both parts of Eq.~(\ref{Gaillard-Zumino-eq}) by 
${\cal P}$ and combine the result with Eq.~(\ref{energy-tensor-conv-nonl}).
After a simple algebra we obtain
\begin{equation}
4\left({\cal S}^2+{\cal P}^2\right){\cal L}_{\cal S}^2
-4\,{\cal L}^2={\cal P}^2\,,
\label
{Gaillard-Zumino-Besse-Hagen}
\end{equation} 
or, equivalently,
\begin{equation}
4\left(\sqrt{{\cal S}^2+{\cal P}^2}\,{\cal L}_{\cal S}-{\cal L}\right)
\left(\sqrt{{\cal S}^2+{\cal P}^2}\,{\cal L}_{\cal S}+{\cal L}\right)={\cal P}^2\,.
\label
{Gaillard-Zumino-Besse-Hagen-}
\end{equation} 
To solve this nonlinear partial differential equation with ${\cal L}$ as the 
unknown function, it is convenient to use 
\begin{equation}
u=\sqrt{{\cal S}^2+{\cal P}^2}\,,
\quad
v={\cal S}\,,
\label
{new-variables}
\end{equation} 
rather than ${\cal S}$ and ${\cal P}$.
These $u$ and $v$ are independent variables everywhere except for the point ${\cal P}=0$ 
which is 
the only singular point of the Gaillard--Zumino condition (\ref{duality-cond}).  
Therefore, we may safely express Eq.~(\ref{Gaillard-Zumino-Besse-Hagen-}) in terms of $u$ and $v$. 
The differentiation with respect to  ${\cal S}$ is 
\begin{equation}
\frac{\partial}{\partial{\cal S}}=\frac{v}{u}\,
\frac{\partial}{\partial u}+\frac{\partial}{\partial v}\,.
\label
{diff-new-variables}
\end{equation} 
Since Eq.~(\ref{energy-tensor-conv-nonl}) is nothing but Euler's homogeneous 
function theorem for homogeneous functions ${\cal L}$ of degree 1, it makes 
sense to look for 
solutions of Eq.~(\ref{Gaillard-Zumino-Besse-Hagen-}) in the form  
\begin{equation}
{\cal L}=\alpha\,u+\beta\,v\,,
\label
{L(u,v)}
\end{equation} 
where $\alpha$ and $\beta$ are unknown constants. 
We substitute the ansatz (\ref{L(u,v)}) into 
Eq.~(\ref{Gaillard-Zumino-Besse-Hagen-}), expressed in terms of $u$ and $v$,
to find
\begin{equation}
4\left(v^2-u^2\right)\left(\alpha^2-\beta^2\right)=\left(u^2-v^2\right).
\label
{Gaillard-Zumino-Besse-Hagen-u-v}
\end{equation} 
It follows that 
\begin{equation}
\alpha =\pm\frac12 \sinh \gamma\,,
\quad
\beta=\pm\frac12 \cosh \gamma\,.
\label
{Gaillard-Zumino-Besse-Hagen-solution}
\end{equation} 
The solution with $\alpha =-\frac12 \sinh \gamma$, $\beta=\frac12 \cosh \gamma$,
$\gamma>0$ represents the Lagrangian which is unbounded from below, and 
should be discarded. 
Hence the desired set of Lagrangians, invariant under conformal group 
transformations and duality rotations (\ref{duality-transf}), is given by the one-parameter 
family of functions 
\begin{equation}
{\cal L}({\cal S},{\cal P};{\gamma})=-\frac12\left({\cal S}\,\cosh \gamma-
\sqrt{{\cal S}^2+{\cal P}^2}\,\sinh \gamma\right),
\label
{final-solution}
\end{equation} 
where the parameter $\gamma$ runs from $0$ to $\infty$, with $\gamma=0$
being attributed to the free Maxwell electrodynamics governed by the 
Larmor Lagrangian ${\cal L}_{\rm L}=-\frac12\,{\cal S}$.

By analogy with Eq.~(\ref{EM-invariants}), we define
\begin{equation}
{\Sigma}=\frac12\,E_{\mu\nu}E^{\mu\nu}\,,
\quad
{\Pi}=\frac12\,E_{\mu\nu}{}^\ast\! E^{\mu\nu}\,.
\label{E-invariants}
\end{equation}
The Gaillard--Zumino criterion (\ref{duality-cond}) requires that ${\Pi}={\cal P}$.
As to the expression for ${\Sigma}$ afforded by the Lagrangian 
(\ref{final-solution}), it is a straightforward matter to establish 
\[
-\frac12\,{\Sigma}=-\frac12\left[\left(\cosh\gamma-\frac{{\cal S}}{\sqrt{{\cal S}^2
+{\cal P}^2}}\,\sinh\gamma \right)F-\frac{{\cal P}}{\sqrt{{\cal S}^2
+{\cal P}^2}}\,\sinh\gamma\,{}^\ast\!F \right]^2
\]
\begin{equation}
=
-\frac12\left({\cal S}\,\cosh 2\gamma-
\sqrt{{\cal S}^2+{\cal P}^2}\,\sinh 2\gamma\right)
={\cal L}({\cal S},{\cal P};{2\gamma})\,,
\label
{EE=L-2gamma}
\end{equation} 
and the inverse 
\begin{equation}
-\frac12\,{\cal S}=
-\frac12\left({\Sigma}\,\cosh 2\gamma-
\sqrt{{\Sigma}^2+{\Pi}^2}\,\sinh 2\gamma\right)
={\cal L}({\Sigma},{\Pi};{2\gamma})\,.
\label
{S=L-2gamma}
\end{equation} 

It is thus seen that the Lagrangian formalism is best suited for a direct and simple 
derivation of Eq.~(\ref{final-solution}). 
On the other hand, the Hamiltonian formalism employed in Ref.~\cite{Bandos}  
may be a convenient framework for analyzing other important problems.
To illustrate, we refer to the fact that the authors of Ref.~\cite{Bandos} were 
fortunate to discover a one-parameter extension of the Born--Infeld theory from 
which Eq.~(\ref{final-solution}) follows immediately in a certain limit.
The Hamiltonian approach makes it clear that the nonlinear extension of the free 
Maxwell electrodynamics unvariant under conformal group transformations and 
duality rotations, embodied in Eq.~(\ref{final-solution}), is unique.
This approach is attractive for the treatment of exact solutions,
specifically solutions of the plane wave type.
Finally, the Hamiltonian  approach is an appropriate starting point for 
exploring quantum properties of the system governed by the Lagrangian
(\ref{final-solution}). 
It is therefore reasonable to invoke the Hamiltonian and Lagrangian approaches 
interchangeably.

I thank Paul Townsend for useful discussions.

\end{document}